\documentclass[10pt,a4paper]{article}

\usepackage{amsfonts}
\usepackage{amssymb}
\usepackage{amsmath}
\usepackage{latexsym}

\usepackage{epsfig}
\usepackage{graphicx}
\usepackage{bm}

\title{\bf \Large  Some remarks on the Einstein and M{\o}ller pseudotensors \\
for static and spherically-symmetric configurations }

\author{Jerzy Matyjasek\footnote{email: jurek@kft.umcs.lublin.pl, matyjase@tytan.umcs.lublin.pl}
\\
\noalign{\vspace{3ex}}
\it{ \small Institute of Physics,  Maria Curie-Sk\l odowska University,}\\
\it{\small pl. Marii Curie-Sk\l odowskiej 1, 20-031 Lublin, Poland}}

\date{}

\begin{document}

\maketitle
\begin{abstract}
It is shown that for the spherically-symmetric and static systems the
hypotheses posed by Yang and Radinschi and by Vagenas can be related to the
particular distribution of the source. Simple proofs are given and a number
of examples are discussed with the special emphasis put on the quantum
corrected Schwarzschild black hole.
\end{abstract}

\vspace{1.5cm}

Recent interest in the pseudotensors describing the energy-momentum of
gravity has been stimulated to a large extent by the understanding that the
freedom in defining the pseudotensors is associated with the freedom in the
choice of the Hamiltonian boundary terms \cite{Nester}. Thus, each
pseudotensor reflects particular physical situation, and, being in fact
quasilocal \cite{york}, is acceptable. This removed (at least partially) the
anathema pronounced by a number of authors (see for example Ref. 3) and
caused that the energy momentum pseudotensors returned from the exile and
have settled themselves in the main stream of the theory. There are quite a
number of various constructions, each of noble parentage: Einstein\cite
{Einstein}, Landau-Lifshitz \cite{LL}, Weinberg \cite{Wein}, M{\o}ller\cite
{Moeller}, Bergman \cite{Berg} and Papapetrou \cite{Pap}, to name a few.
Each has been a subject of detailed studies and a source of interesting and
important results. The subjects covered range form the black hole physics 
to gravitational waves. (See, for example, Refs. 10-18 and
references cited therein). Despite extensive work done so far there is still
room for new findings and fresh insights.

Some time ago Radinschi and Yang  \cite{Irina}\ observed that for the
Reissner-Nordstr\"{o}m and the regular black holes \cite
{ABG,Bronnikov,kocio,IrinaD} the difference between the energy calculated
with the aid of the Einstein pseudotensor, $E_{E},$ and this obtained
employing the M{\o}ller prescription, $E_{M},$ can be related to the energy
density, $\rho $, of the matter fields 
\begin{equation}
E_{E}-E_{M}\sim \text{ }r^{3}\rho .  \label{irin_eq}
\end{equation}
Later Vagenas \cite{Elias}\ found that for the Reissner-Nordstr\"{o}m (RN)
and the RN dyadosphere \cite{Remo} there is an interesting relation between
the coefficients of the expansion of $E_{E}$ and $E_{M}$ in the inverse
powers of $r.$  This has been partially confirmed by the behaviour of a few
first coefficients of the expansion of the regular black hole solution. He
hypothesized that if $\alpha _{k}$ and $\beta _{k}$ are, respectively, the
expansion coefficients of $E_{E}$ and $E_{M}$, then 
\begin{equation}
\alpha _{k}=\frac{1}{k+1}\beta _{k}.  \label{vag_eq}
\end{equation}

It is the aim of this letter to demonstrate how these observations and
hypotheses can be proven and understood in relation with the source term of
the Einstein field equations, (i. e., the stress-energy tensor of the matter
fields) and to analyse their limitations. The discussion will be illustrated
with a few examples. As our demonstration heavily relies on the particular
representation of the line element describing static and
spherically-symmetric configuration we shall start with a few basic facts.

Let us consider the general spherically-symmetric, static and asymptotically
flat geometry described by the line element (our conventions follow these of
MTW) 
\begin{equation}
ds^{2}=-e^{2\psi \left( r\right) }\left( 1-\frac{2M(r)}{r}\right)
dt^{2}+\left( 1-\frac{2M(r)}{r}\right) ^{-1}dr^{2}+r^{2}\left( d\theta
^{2}+\sin ^{2}\theta d\phi ^{2}\right) ,  \label{gen_el}
\end{equation}
where $M\left( r\right) $ and $\psi \left( r\right) $ are two functions that
are to be determined from the Einstein field equations, supplemented by the
physically motivated boundary conditions. This representation of the line
element has proven to be useful in various contexts as the Einstein field
equations have particularly transparent form \cite{visser}. Moreover, let us
assume that the (total) covariantly conserved stress-energy tensor of the
material fields can be written in the form dictated by the symmetries of the
problem 
\begin{equation}
T_{\nu }^{\mu }=\mathrm{diag}\left[ a\left( r\right) ,b\left( r\right)
,c\left( r\right) ,c\left( r\right) \right] .  \label{Stress}
\end{equation}
The $\left( _{0}^{0}\right) $ component of the Einstein equations reads 
\begin{equation}
-\frac{2}{r^{2}}\frac{dM\left( r\right) }{dr}=8\pi a\left( r\right) ,
\label{row1}
\end{equation}
whereas their $\left( _{1}^{1}\right) $ component can be written as 
\begin{equation}
-\frac{2}{r^{2}}\frac{dM\left( r\right) }{dr}+\frac{2}{r}\left( 1-\frac{%
2M\left( r\right) }{r}\right) \psi ^{\prime }\left( r\right) =8\pi b\left(
r\right) .  \label{row2}
\end{equation}
In the latter we shall employ the boundary conditions in the form 
\begin{equation}
M\left( \infty \right) =\mathcal{M,\hspace{0.75cm}}\psi \left( \infty
\right) =0,  \label{b1}
\end{equation}
where $\mathcal{M}$ is the total mass of the system as seen by a distant
observer. On the other hand, for the black hole configurations, one can also
take 
\begin{equation}
M\left( r_{+}\right) =\frac{r_{+}}{2}\mathcal{,\hspace{0.75cm}}\psi \left(
\infty \right) =0,  \label{b2}
\end{equation}
where $r_{+}$ is the location of the event horizon, defined here as the
outermost trapped surface. \ From (\ref{row1}) one has 
\begin{equation}
\frac{dM\left( r\right) }{dr}=-4\pi r^{2}a\left( r\right) ,  \label{eq1}
\end{equation}
whereas from (\ref{row1}) and (\ref{row2}) one obtains 
\begin{equation}
\frac{d\psi \left( r\right) }{dr}=4\pi r\frac{b\left( r\right) -a\left(
r\right) }{1-\frac{2M\left( r\right) }{r}}.  \label{eq2}
\end{equation}

Now, let us return to the general form of the line element and calculate the
Einstein pseudotensor \cite{Einstein}. The Einstein pseudotensor, $\Theta
_{\lambda }^{\mu },$ can easily be constructed in the Cartesian-type
coordinates by differentiation of the Freud superpotential, $V_{\lambda
}^{\mu \nu },$%
\begin{equation}
\Theta _{\lambda }^{\mu }=\frac{1}{16\pi }\frac{\partial }{\partial x^{\nu }}%
V_{\lambda }^{\mu \nu },  \label{ein}
\end{equation}
where 
\begin{equation}
V_{\lambda }^{\mu \nu }=\frac{1}{\sqrt{-g}}g_{\lambda \tau }\frac{\partial }{%
\partial x^{\gamma }}\left[ g\left( g^{\mu \tau }g^{\nu \gamma }-g^{\nu \tau
}g^{\mu \gamma }\right) \right] .  \label{sup_eins}
\end{equation}

The $\Theta _{\lambda }^{\mu }$ does not exhaust the list of interesting and
important pseudotensors. Indeed, there are half a dozen other objects
(sometime referred to as complexes) that are useful in various
circumstances. One of the most studied, however, is the pseudotensor
constructed by M{\o}ller in the late fifties of the last century. One of
the advantages of the M{\o}ller pseudotensor, $\mathcal{T}_{\nu }^{\mu },$
is that it can be calculated in any coordinate system. Its construction
proceeds in two steps: first, one has to calculate the superpotential 
\begin{equation}
U_{\nu }^{\mu \lambda }=-\sqrt{-g}\left( \frac{\partial }{\partial x^{\kappa
}}g_{\nu \sigma }-\frac{\partial }{\partial x^{\sigma }}g_{\nu \kappa
}\right) g^{\mu \kappa }g^{\lambda \sigma },  \label{sup_mel}
\end{equation}
and, subsequently, the energy-momentum pseudotensor 
\begin{equation}
\mathcal{T}_{\nu }^{\mu }=\frac{1}{8\pi }\frac{\partial }{\partial
x^{\lambda }}U_{\nu }^{\mu \lambda }.  \label{mel}
\end{equation}

Due to the antisymmetry of the superpotentials the pseudotensors (\ref{ein})
and (\ref{mel}) are divergence-free. The energy and momentum are given by 
\begin{equation}
P_{i}=\iiint\mathcal{\tau}_{i}^{0}dx^{1}dx^{2}dx^{3}  \label{momentum}
\end{equation}
and 
\begin{equation}
E=\iiint\mathcal{\tau}_{0}^{0}dx^{1}dx^{2}dx^{3},  \label{energy}
\end{equation}
respectively, where $\mathcal{\tau}_{\nu}^{\mu}$ stands either for $%
\Theta_{\nu}^{\mu}$ or $\mathcal{T}_{\nu}^{\mu}.$ In the latter we shall
differentiate between them simply by inserting subscripts $E$ or $M.$

The energy calculated with the aid of the M{\o}ller energy-momentum
pseudotensor for the geometry (\ref{gen_el}) assumes the simple form 
\begin{equation}
E_{M}=e^{\psi \left( r\right) }\left[ M\left( r\right) -rM^{\prime }\left(
r\right) +r^{2}\psi ^{\prime }\left( r\right) -2rM\left( r\right) \psi
^{\prime }\left( r\right) \right] .  \label{Moller}
\end{equation}
On the other hand, the calculations of the Einstein pseudotensor is slightly
more involved. Indeed, as have been said earlier one has to introduce the
standard quasi-Cartesian coordinates $\{x,y,z\}$ and rewrite the line
element (\ref{gen_el}) in the form 
\begin{align}
ds^{2}& =-e^{2\psi \left( r\right) }\left( 1-\frac{2M(r)}{r}\right)
dt^{2}+\left( dx^{2}+dy^{2}+dz^{2}\right)   \notag \\
& +\frac{1}{r^{2}}\left[ \left( 1-\frac{2M(r)}{r}\right) ^{-1}-1\right]
\left( xdx+ydy+zdz\right) ^{2}.  \label{quasi}
\end{align}
From the definition of the Einstein pseudotensor, after some algebra, one
obtains remarkably simple result 
\begin{equation}
E_{E}=e^{\psi \left( r\right) }M\left( r\right) .  \label{Ee}
\end{equation}
The calculations have been carried out with the aid of Maxima package, and
in the advent of the computer algebra systems, this problem should be
regarded as a five-finger exercise. Identical results can easily be obtained
from the formulas constructed by Virbhadra \cite{Virb1} and Xulu \cite{Xulu}%
\ for the Einstein and M{\o}ller pseudotensors, respectively, in the most
general nonstatic and spherically-symmetric geometry. 

Now, let us assume that $a\left( r\right) $ falls sufficiently rapidly as $%
r\rightarrow \infty $ to make the integral (\ref{eq1}) finite and $a\left(
r\right) =b\left( r\right). $ Making use of the boundary conditions (\ref
{b1}) one has $\psi \left( r\right) =0$ and 
\begin{equation}
M\left( r\right) =\mathcal{M}+4\pi \int_{\infty }^{r}r^{2}a\left( r\right)
dr=\mathcal{M}+M_{0}\left( r\right) .
                                    \label{eq20}
\end{equation}
On the other hand, the boundary conditions (\ref{b2}) yield 
\begin{equation}
M\left( r\right) =\frac{r_{+}}{2}+4\pi \int_{r_{+}}^{r}r^{2}a\left( r\right)
dr=\frac{r_{+}}{2}+\tilde{M}_{0}\left( r\right) .
\end{equation}

For the simplified configuration the M{\o}ller energy reads 
\begin{equation}
E_{M}=M\left( r\right) -rM^{\prime }\left( r\right) ,  \label{e_mel}
\end{equation}
whereas the Einstein energy coincides with the function $M\left( r\right) $
as can  easily be seen from Eq.(\ref{Ee}): 
\begin{equation}
E_{E}=M\left( r\right) .  \label{e_eins}
\end{equation}

Now, we are ready to demonstrate a simple relation existing between the
expansions (in terms of $r^{-1}$) of the Einstein and M{\o}ller energy.
Following Vagenas, let us assume that 
\begin{equation}
M_{0}\left( r\right) =\sum_{k=1}^{\infty }\alpha _{k}r^{-k},  \label{m0E}
\end{equation}
where $\alpha _{k\text{ }}$are numerical coefficients. It should be noted
that there are quite a number of solutions that belong to that class.
Indeed, the Reissner-Nordstr\"{o}m, the regular black holes or the
geometry of the dyadosphere of RN black hole may serve as the simple
examples of particular realizations of (\ref{m0E}) with 
\begin{equation}
\alpha _{1}=-\frac{Q^{2}}{2}\text{, \ \ }\alpha _{k}=0\text{ \ \ for \ }%
k\geq 2,  \label{rn}
\end{equation}
\begin{equation}
\alpha _{k}=\frac{1}{k!}\frac{d^{k}}{d\xi ^{k}}M_{0}\left( \xi \right)
_{|\xi =0,}\text{ \ \ \ \ \ }\xi =\frac{1}{r},  \label{abg}
\end{equation}
and 
\begin{equation}
\alpha _{1}=-\frac{Q^{2}}{2},\text{ \ }a_{2}=\alpha _{3}=\alpha _{4}=0,\text{
\ }\alpha _{5}=\sigma \frac{Q^{4}}{10},\text{ \ }\alpha _{k}=0\text{ \ \ for
\ }k\geq 6,
\end{equation}
respectively. We do not attempt, of course, to list all the solutions of
this type here.

It can easily be demonstrated that if $M_{0}\left( r\right) $ is given by (%
\ref{m0E}) then 
\begin{equation}
E_{E}=\mathcal{M}+\sum_{k=1}^{\infty }\alpha _{k}r^{-k}
\end{equation}
and 
\begin{equation}
E_{M}=\mathcal{M}+\sum_{k=1}^{\infty }\left( k+1\right) \alpha
_{k}r^{-k}\equiv \mathcal{M}+\sum_{k=1}^{\infty }\beta _{k}r^{-k},
\label{mm}
\end{equation}
and, consequently, 
\begin{equation}
\alpha _{k}=\frac{1}{k+1}\beta _{k}.  \label{Elias}
\end{equation}
The above equation is precisely the relations found by Vagenas \cite{Elias}.
He correctly hypothesized that it should have a wider domain of
applicability than two or three special cases considered in Ref. 18. It
should be emphasized, however, that in this demonstration the crucial
role is played by the special form of the stress-energy tensor. 

On the other hand, still working with the simplified \ stress-energy tensors
of the matter fields one has 
\begin{align}
\Delta E& =E_{M}-E_{E}=-rM^{\prime }\left( r\right)  \notag \\
& =4\pi r^{3}a(r)=4\pi r^{3}T_{0}^{0},  \label{Radinschi}
\end{align}
where we have used Eq. (\ref{eq1}). This result for particular
configurations has been obtained by Yang and Radinschi in Ref. 17.
Finally observe, that for $a\left( r\right) \neq b\left( r\right) $ the
relations (\ref{Elias})  and (\ref{Radinschi}) do not hold in general.

Equipped with the results (\ref{e_mel}) and (\ref{e_eins}) let us consider a
slightly more complicated example of the quantum corrected black hole being
a solution of the semiclassical Einstein field equations. 
The idea that lies behind the calculations is simple. One has to
construct the stress-energy tensor for a general background (\ref{gen_el})
and subsequently solve (self-consistently) the differential
equations with the boundary conditions (\ref{b1}) or (\ref{b2}).

The stress-energy
tensor of the quantized massive scalar, spinor and vector fields can be
obtained from the one-loop renormalized effective action, $W_{R},$ in a standard way,
i.e., by a functional differentiation of $W_{R}$ with respect to the metric
tensor. It gives rise to the quantum-corrected black hole described by the line
element~(\ref{gen_el}). The most general expression describing the approximate
renormalized stress-energy tensor of the quantized massive scalar, spinor and
vector fields have been constructed in Refs. 27 and 28.
These formulas generalize the earlier results obtained by Frolov and 
Zel'nikov  \cite{FZ1,FZ2,FZ3} (valid for the Ricci-flat geometries) and those of  
Ref. 32, where the stress-energy tensor has been calculated for the
massive scalar fields in the general static, spherically-symmetric
geometries. 

In what follows we shall employ
the boundary conditions (\ref{b1}), i.e., we shall parametrize the solution by the
total mass of the system (consisitng of both classical and quantum parts)
rather than the bare black hole mass. The resulting differential
equations are, in general, too complicated to be solved exactly and
consequently one has to refer to approximations. Fortunately, one can easily
adopt an approach in which the stress-energy tensor of the quantized fields
can be regarded as a small perturbation. Indeed, this can be done simply
because the total stress-energy tensor of the matter fields can be written
as 
\begin{equation}
T_{\nu }^{\mu }=\mathrm{diag}\left[ a\left( r\right) +\varepsilon \langle
T_{t}^{t}\rangle _{ren}^{\left( s\right) },b\left( r\right) +\varepsilon
\langle T_{r}^{r}\rangle _{ren}^{\left( s\right) },c\left( r\right)
+\varepsilon \langle T_{\theta }^{\theta }\rangle _{ren}^{\left( s\right)
},c\left( r\right) +\varepsilon \langle T_{\phi }^{\phi }\rangle _{ren}^{\left(
s\right) }\right] ,  \label{t_quant}
\end{equation}
where $\langle T_{\nu }^{\mu }\rangle _{ren}^{\left( s\right) }$ is the
approximate renormalized stress-energy tensor of the quantized massive
scalar ($s=0)$, spinor ($s=1/2$) and vector ($s=1$) fields, and $\varepsilon $
is the auxiliary parameter that helps to keep track of various terms in
complicated expansions. (One should put $\varepsilon =1$ in a final stage of
calculations). 

To simplify discussion let us assume that for the classical
sources the condition $a\left( r\right) =b\left( r\right) $ holds. This
allows us to represent the functions $M\left( r\right) $ and $\psi \left(
r\right) ,$ with a small abuse of notation, in the form\footnote{The function 
$M_{0}\left( r\right) $ in Eq.~\ref{empsi} must not be  confused with the one defined
in Eq.~~\ref{eq20}. }
\begin{equation}
M\left( r\right) =M_{0}\left( r\right) +\varepsilon M_{1}\left( r\right)
+O\left( \varepsilon ^{2}\right) \text{ \ \ \ \ and \ \ \ \ }\psi \left(
r\right) =\varepsilon \psi _{1}\left( r\right) +O\left( \varepsilon ^{2}\right) 
\text{ .}  \label{empsi}
\end{equation}
Now, restricting to the first-order perturbations one has
the following system of the differential equations for $M_{1}\left( r\right) 
$ and $\psi _{1}\left( r\right) $: 
\begin{equation}
\frac{dM_{1}\left( r\right) }{dr}=-4\pi r^{2}\langle T_{t}^{t}\rangle
_{ren}^{\left( s\right) }  \label{q_1}
\end{equation}
and 
\begin{equation}
\frac{d\psi _{1}\left( r\right) }{dr}=4\pi r\frac{\langle T_{r}^{r}\rangle
_{ren}^{\left( s\right) }-\langle T_{t}^{t}\rangle _{ren}^{\left( s\right) }%
}{1-\frac{2M_{0}\left( r\right) }{r}}.  \label{q_2}
\end{equation}
The energy calculated from the Einstein pseudotensor is 
\begin{equation}
E_{E}=M_{0}\left( r\right) +\left[ M_{0}\left( r\right) \psi _{1}\left(
r\right) +M_{1}^{\prime }\left( r\right) \right] \varepsilon +O\left( \varepsilon
^{2}\right)   \label{q_E}
\end{equation}
whereas the M{\o}ller energy is given by 
\begin{align}
E_{M}& =M_{0}\left( r\right) -rM_{0}^{\prime }\left( r\right) +\left[
M_{0}\left( r\right) \psi _{1}\left( r\right) +M_{1}^{\prime }\left(
r\right) -r\psi _{1}\left( r\right) M_{0}^{\prime }\right.   \notag \\
& \left. -rM_{1}^{\prime }\left( r\right) +r^{2}\psi _{1}^{\prime }\left(
r\right) -2rM_{0}\left( r\right) \psi _{1}^{\prime }\left( r\right) \right]
\varepsilon +O\left( \varepsilon ^{2}\right) ;  \label{q_M}
\end{align}
\ and the difference $\Delta E$ is 
\begin{equation}
\Delta E=-rM_{0}^{\prime }\left( r\right) -\left[ \psi _{1}\left( r\right)
M_{0}^{\prime }+rM_{1}^{\prime }\left( r\right) -r^{2}\psi _{1}^{\prime
}\left( r\right) +2rM_{0}\left( r\right) \psi _{1}^{\prime }\left( r\right) %
\right] \varepsilon +O\left( \varepsilon ^{2}\right) .  \label{q_d}
\end{equation}

In the simplest case, the classical sources are absent and the zeroth-order
solution reduces to the Schwarzschild line element with $M_{0}\left(
r\right) =\mathcal{M}.$ (Note that \ $M_{0}\left( r\right) $ $=\mathcal{M}$
and $M_{1}\left( r\right) $ $=0$ as $r\rightarrow \infty $).  Although the
renormalized stress-energy tensor in the Schwarzschild spacetime has been
calculated by Frolov and Zel'nikov \cite{FZ1} in the present calculations we have used
general formulas of  Refs. 27 and 28. After some algebra, for
the quantized massive scalar field ($s=0)$ in a large mass limit with
arbitrary curvature coupling, one has 
\begin{equation}
M^{\left( 0\right) }\left( r\right) =\mathcal{M}-\frac{\mathcal{M}^{2}}{%
15\pi m^{2}r^{6}}\left[ \left( 11\mathcal{M}-6r\right) {\eta }+{\frac{19}{56}%
}r-{\frac{313}{504}}\,\mathcal{M}\right]  \label{first}
\end{equation}
and 
\begin{equation}
\psi ^{\left( 0\right) }\left( r\right) =\frac{\mathcal{M}^{2}}{\pi
m^{2}r^{6}}\left( \frac{7}{15}\eta -\frac{13}{504}\right) .
\end{equation}
Here $m$ is a mass of the field whereas $\eta =\xi -1/6$ is the numerical
parameter and its two particularly appealing values are $\xi =0$ for the
minimal and $\xi =1/6$ for the conformal coupling. Similarly, for the
quantized massive spinor the first-order solution is simply given by 
\begin{equation}
M^{\left( 1/2\right) }\left( r\right) =\mathcal{M}-\frac{\mathcal{M}^{2}}{%
140\pi m^{2}r^{6}}\left( 3r-\frac{149}{27}\mathcal{M}\right) ,
\end{equation}
\begin{equation}
\psi ^{\left( 1/2\right) }\left( r\right) =-\frac{11\mathcal{M}^{2}}{420\pi
m^{2}r^{6}}
\end{equation}
whereas for the quantized massive vector field one has 
\begin{equation}
M^{\left( 1\right) }\left( r\right) =\mathcal{M}+\frac{\mathcal{M}^{2}}{%
280\pi m^{2}r^{6}}\left( 37r-\frac{611}{9}\mathcal{M}\right) ,
\end{equation}
\begin{equation}
\psi ^{\left( 1\right) }\left( r\right) =\frac{131\mathcal{M}^{2}}{840\pi
m^{2}r^{6}}.  \label{last}
\end{equation}
Eqs.(\ref{q_E}), (\ref{q_M}) and (\ref{first}-\ref{last}) are, of course,
sufficient to calculate the Einstein and M{\o}ller energy. $\,$On the other
hand, combining (\ref{q_d}), \ (\ref{q_1}) \ and (\ref{q_2}), for a problem
on hand,  for $\Delta E$ one has
\begin{equation}
\Delta E=4\pi r^{3}\langle T_{r}^{r}\rangle _{ren}^{\left( s\right) }.
\label{qq}
\end{equation}
It is an interesting result, which clearly shows that the difference between
the M{\o}ller and Einstein energy is related to the radial pressure of the
quantized fields. It can be contrasted with the Radinschi-Yang formula.
One can ascribe this behaviour to $\psi \left( r\right) $ $
\neq 0$, which, in turn, is a consequence of the form of the stress-energy
tensor. On the other hand, for the class of the renormalized stress-energy
tensors of the quantized fields in a large mass limit in the
Bertotti-Robinson (or more generally $AdS_{2}\times S_{2}$) geometries one
has $\psi \left( r\right) =0$. It should be noted that the formula (\ref{qq}%
) may be used as the useful check of the calculations. 

Now, making use of either (\ref{q_d}) or (\ref{qq}), one readily obtains 
\begin{equation}
\Delta E^{\left( 0\right) }=\frac{\mathcal{M}^{2}}{\pi m^{2}r^{6}}\left[ 
\frac{2}{5}\left( 3\mathcal{M-}2r\right) \eta +\frac{1}{24}r-\frac{11}{180}%
\mathcal{M}\right] =4\pi r^{3}\langle T_{r}^{r}\rangle _{ren}^{\left(
0\right) },  \label{dd1}
\end{equation}

\begin{equation}
\Delta E^{\left( 1/2\right) }=\frac{\mathcal{M}^{2}}{\pi m^{2}r^{6}}\left( 
\frac{r}{20}-\frac{7}{90}\mathcal{M}\right) =4\pi r^{3}\langle
T_{r}^{r}\rangle _{ren}^{\left( 1/2\right) }  \label{dd2}
\end{equation}
and 
\begin{equation}
\Delta E^{\left( 1\right) }=-\frac{\mathcal{M}^{2}}{\pi m^{2}r^{6}}\left( 
\frac{11r}{40}-\frac{5}{12}\mathcal{M}\right) =4\pi r^{3}\langle
T_{r}^{r}\rangle _{ren}^{\left( 1\right) }.  \label{dd3}
\end{equation}
It should be emphasized once more, that $\langle T_{\mu }^{\nu }\rangle
_{ren}^{\left( s\right) }$ calculated for the zeroth-order solution of the
semiclassical Einstein field equations coincides with the stress energy
tensor of the massive fields constructed in the Schwarzschild spacetime
parametrized by the mass $\mathcal{M.}$

To this end, let us calculate the Einstein energy as the M{\o}ller energy
can easily be obtained from the definition of $\Delta E^{\left( s\right) }$
and (\ref{dd1}-\ref{dd3}). Inserting \ the functions $M^{\left( s\right)
}\left( r\right) $ and $\psi ^{\left( s\right) }\left( r\right) $ as given
by (\ref{first}-\ref{last}) into (\ref{q_E}) (or, equivalently, into (\ref
{Ee}) and retaining the terms which are linear in $\varepsilon $) one obtains 
\begin{equation}
E_{E}^{\left( 0\right) }=\mathcal{M}+\frac{\mathcal{M}^{2}}{15\pi m^{2}r^{6}}%
\left[ \frac{59}{252}\mathcal{M}-\frac{19}{56}r+\left( 6r-4\mathcal{M}%
\right) \eta \right] ,  \label{eee}
\end{equation}
\begin{equation}
E_{E}^{\left( 1/2\right) }=\mathcal{M}+\frac{\mathcal{M}^{2}}{\pi m^{2}r^{6}}%
\left( \frac{5}{378}\mathcal{M}-\frac{3}{140}r\right)  \label{ee1}
\end{equation}
and 
\begin{equation}
E_{E}^{\left( 1\right) }=\mathcal{M}-\frac{\mathcal{M}^{2}}{280\pi m^{2}r^{6}%
}\left( \frac{218}{9}\mathcal{M}-37r\right) .  \label{ee2}
\end{equation}

Finally let us return to the boundary conditions (\ref{b2}) and observe that
in this case the solution is parametrized by the exact location of the event
horizon, $r_{+}$. The total mass of the system is, therefore, given by 
\begin{equation}
\mathcal{M}=\lim_{r\rightarrow \infty }\left( M_{0\left( r\right) }+\varepsilon
M_{1}\left( r\right) \right) ,
\end{equation}
and it differs from the horizon defined mass, $M_{H}$. As the
calculations in this parametrization closely follow the general pattern
described above, we shall not display them here.

Our next example is to certain extent related to the previous one: we shall
consider the pure gravity action supplemented with the higher curvature
terms and analyse the Schwarzschild solutions influenced by all
time-reversal-invariant operators of dimension six \cite{LuWise}. In doing
so we shall assume that the coefficients of the higher curvature part of the
action are arbitrary rather than inspired by a particular theory. The action
functional has, therefore, the form 
\begin{equation}
S=\int d^{4}xg^{1/2}R+S_{1}+S_{2},
\end{equation}
where $S_{1}$ is the action functional of the quadratic gravity (that does
not influence the problem) and $S_{2}$ is given by

\begin{eqnarray}
S_{2} &=&\int d^{4}xg^{1/2}\left( c_{1}R_{\mu \nu \lambda \sigma }R^{\alpha
\beta \lambda \sigma }R_{~~~\alpha \beta }^{\mu \nu }+c_{2}R_{~~~\lambda
\sigma }^{\mu \nu }R_{\mu \alpha }^{~~~~\lambda \beta }R_{~~~\nu \beta
}^{\alpha \sigma }\right.  \notag \\
&&+c_{3}R_{\mu \nu }R^{\mu \alpha \beta \gamma }R_{~~\alpha \beta \gamma
}^{\nu }+c_{4}RR_{\mu \nu \lambda \kappa }R^{\mu \nu \lambda \kappa }  \notag
\\
&&+\left. c_{5}R_{\mu \nu }R_{\lambda \kappa }R^{\mu \lambda \nu \kappa
}+c_{6}R_{\mu \nu }R^{\mu \lambda }R_{\lambda }^{~~\nu
}+c_{7}R^{3}+c_{8}RR_{\mu \nu }R^{\mu \nu }\right) \,\,.  \label{high}
\end{eqnarray}
For spherically-symmetric and static systems the perturbative solution is
given by (\ref{gen_el}) with

\begin{equation}
\psi (r)=-\frac{96\pi }{r^{6}}\mathcal{M}^{2}\varepsilon \left(
18 c_{1}+6 c_{2}+ 7 1c_{3}+ 28 c_{4}\right)   \label{pso_lw}
\end{equation}
and 
\begin{eqnarray}
M\left( r\right)  &=&\mathcal{M}-\frac{144\pi \mathcal{M}^{2}}{r^{5}}
\varepsilon \left( 12c_{1}+3c_{2}+4c_{3}+16c_{4}\right) \,  \notag \\
&&+{\frac{32\pi \mathcal{M}^{3}}{r^{6}}}\,\varepsilon \left(
98c_{1}+25c_{2}+33c_{3}+132c_{4}\right) .  \label{M_lw}
\end{eqnarray}
Elementary calculation shows that the difference between Einstein
and M{\o}ller complexes is given by
\begin{equation}
\Delta E=-rM_{1}^{\prime }\left( r\right) +r^{2}\psi _{1}^{\prime }\left(
r\right) \left( 1-\frac{2\mathcal{M}}{r}\right).  
\label{ps_LW}
\end{equation}
Making use of  (\ref{M_lw}) and (\ref{ps_LW}), one  arrives at 
\begin{eqnarray}
\Delta E &=&\frac{144\pi \mathcal{M}^{2}}{r^{5}}\left(
16 c_{1}+9 c_{2}+8 c_{3}+ 32c_{4}\right)   \notag \\
&&-\frac{192\pi \mathcal{M}^{3}}{r^{6}}\left(
10 c_{1}+11 c_{2}+9 c_{3}+36 c_{4}\right) .
\end{eqnarray}
It can easily be demonstrated by an independent calculation that 
\begin{equation}
S_{r}^{r}=\Delta E/8\pi r^{3},
\end{equation}
where 
\begin{equation}
\mathcal{S}^{ab}=\frac{1}{g^{1/2}}\frac{\delta S_{2}}{\delta g_{ab}}.
\end{equation}
Treating $S^{ab}$ as the effective stress-energy tensor, one concludes
that $\Delta E$ is proportional to the radial pressure, as expected.

The results obtained in this paper clearly shows that it is necessary to
explain why various pseudotensors are related in a specific way. We suspect
that there are other, quite simple but interesting relations which can
easily be explained with the aid of the Einstein equations and careful
examination of the role played by their source terms. Moreover, it would be
interesting to consider other pseudotensors for the whole classes of
geometries and relate them with the specific choices of the boundary terms.
Finally we observe that the evident direction for further research is
generalization of the semi-classical calculations to more complicated
settings, such as the Reissner-Nordstr\"{o}m spacetime or
the geometry of the regular black holes. These issues are under 
active consideration and the
results will be published elsewhere.

\section*{Acknowledgments}

The author acknowledges helpful correspondence with Elias Vagenas and his
critical reading of the manuscript.


\end{document}